\documentclass[aps,pre,twocolumn,reprint,superscriptaddress,longbibliography,floatfix]{revtex4-2}

\usepackage{amsmath,amssymb,bm,mathtools}
\usepackage{graphicx}
\usepackage[dvipsnames]{xcolor}
\usepackage{xspace}
\usepackage[breaklinks,colorlinks,citecolor=MidnightBlue,linkcolor=MidnightBlue,urlcolor=MidnightBlue]{hyperref}

\begin{document}

\title{Reentrance in a Hamiltonian flocking model}

\author{Letian Chen}
\affiliation{School of Mathematics and Maxwell Institute for Mathematical Sciences, University of Edinburgh, Edinburgh, EH9 3FD, Scotland}

\author{Luke K. Davis}
\email{luke.davis@ed.ac.uk}
\affiliation{School of Mathematics and Maxwell Institute for Mathematical Sciences, University of Edinburgh, Edinburgh, EH9 3FD, Scotland}
\affiliation{Higgs Centre for Theoretical Physics, University of Edinburgh, Edinburgh, EH9 3FD, Scotland}

\date{\today}

\begin{abstract}

The clustering of self-motile and repulsive particles, so-called motility-induced phase separation (MIPS), is one of the clearest signatures of active physics. Typically, increasing the amplitude of self-motility increases the degree of clustering, however for high enough self-motility the homogeneous phase is reentered. 
Here, we report that such reentrance naturally emerges in a Hamiltonian (conservative) model known to recapitulate properties of (active) bird flocks, and exhibits clustering behaviour reminiscent of MIPS. We numerically demonstrate the reentrance of the homogeneous phase and identify the underlying mechanism as a competition between the amplitude of a spin-velocity coupled drive and mobility-limited kinetic frustration.  Specifically, we reveal that strong spin-velocity coupling suppresses transverse diffusion, thereby leading the system into an arrest that closes the window for phase separation. Overall, our work offers a Hamiltonian, conservative, bridge between reentrant physics across equilibrium and non-equilibrium materials.

\end{abstract}

\keywords{}

\maketitle
\section{Introduction}

Reentrance is where a system transitions from one phase to another phase, and back to the original phase upon the monotonic driving of a physical parameter. Reentrance is a ubiquitous phenomenon in soft and biological matter, with examples ranging from multi-component colloidal fluids \cite{narayanan_1994,likos_2001} and protein assemblies \cite{Zhang_2008} to patchy particle systems \cite{Russo_2011}. Reentrance typically brings about non-monotonicity in physical observables which resume similar values of the original phase. Such non-monotonic behavior has also garnered significant attention in active matter \cite{Mingfeng_2017,Arora_2022,Paoluzzi2022,Berx_2023,Evans2024,Paoluzzi2024,Baconnier_2025,desouza_2025,Puneet_2026,Burekovic2026}. For active systems consisting of self-propelled and repulsive particles, at an appropriate density, increasing the P\'{e}clet number (Pe) takes the system from a homogeneous phase to one exhibiting motility-induced phase separation (MIPS) \cite{Tailleur2008,Fily_2012,Redner_2013_2, Cates2015,Stenhammar2015,Digregorio2018,Klamser_2018}, with reports that further increases in Pe resulting in reentrance of the homogeneous phase \cite{Paoluzzi2020,Su2023,Evans2024,Yamamoto2025}. 

This phenomenon challenges our understanding of how microscopic driving dictates macroscopic behavior and, while reentrance generally implies underlying frustration \cite{Thomas2011}, or competing interactions \cite{Reinhardt2011}, the precise microscopic mechanisms governing the reentrance in active systems have yet to be fully elucidated. A fundamental open question is whether the type of reentrance seen in active systems is exclusive to systems with non-conservative energy injection, or if it can also arise in conservative systems.

Hamiltonian flocking models (HFMs) offer an ideal testing ground for this inquiry \cite{LolandBore2016,Casiulis_2019,Casiulis2020,Bhattacharya2025}. By incorporating spin-velocity, as well as standard ferromagnetic, coupling into a conservative framework, HFMs recapitulate aspects of flocking and MIPS without explicit self-propulsion or attractive interactions. Without the spin-velocity term, and only with repulsion and spin-spin coupling, the system can undergo ferromagnetic-induced phase separation (FIPS). Unlike standard active matter models where activity is a hard-coded non-conservative parameter introduced on the level of the dynamics, HFMs generate effective drive through spin-velocity couplings introduced on the Hamiltonian/Lagrangian level, allowing for a rigorous examination of phase behavior within a thermodynamically consistent dynamical framework.

Here, we numerically demonstrate that reentrance in (repulsive) phase separating systems is not unique to non-conservative active matter models, but naturally emerges in a HFM. We systematically show that reentrance is driven by a competition between coupling-induced effective drive and mobility-limited kinetic frustration. Specifically, we reveal that strong spin-velocity coupling leads to an effective dimensional reduction that suppresses transverse diffusion, thereby constraining the particle rearrangements necessary for cluster formation. Based on this mechanism, we arrive at a parameter-free scaling ansatz that well captures the non-monotonic phase boundaries observed in the simulations.

\section{Model}
In the HFM the Hamiltonian is written as follows \cite{Casiulis2020,Bhattacharya2025}:
\begin{equation}\label{eq:Hamilton}
   \begin{aligned}
\mathcal{H}&:= \sum_{i=1}^N \left( \frac{({\mathbf{p}}_i - K \mathbf{S}_i)^2}{2m} + \frac{\omega_i^2}{2I} \right)
 \\
 &\qquad \qquad + \sum_{i<j}^N \bigg( U_\mathrm{R}(r_{ij}) - g(r_{ij})\mathbf{S}_i\cdot\mathbf{S}_j \bigg),
 \end{aligned}
\end{equation}
where $N$ is the number of particles, $m$ is the mass of each particle, $\mathbf{p}_i$ is the linear momentum of particle $i$, $I$ is the moment of inertia of each particle, $\omega_i$ is the per-particle angular momentum, $K$ is the spin-velocity coupling, $\mathbf{S}_i$ is the spin on particle $i$, $r_{ij} \equiv |\mathbf{r}_i - \mathbf{r}_j|$ is the separation between $i$ and $j$. $U_\text{R}(r)$ is a repulsive interaction given by the Weeks-Chandler-Anderson (WCA) pair potential:
\begin{equation}
    U_\text{R}(r) = \left(4\varepsilon \left[ \left( \frac{\sigma}{r}\right)^{12}- \left(\frac{\sigma}{r}\right)^{6} \right] + \varepsilon\right) \Theta(2^{\frac{1}{6}} \sigma -r),
\end{equation}
where $\sigma=1$ is the range of the interaction, $\varepsilon  = 10$ is the interaction strength, whose value is so chosen to ensure only minor overlapping of particles, and $\Theta(\cdots)$ is the Heaviside function. The ferromagnetic spin-spin interaction range is smoothly truncated using a step function kernel $g(r)$:
\begin{equation}\label{eq:step_kernel}
    g(r) = J \frac{(r-\frac{3}{2} \sigma)^4}{\epsilon + (r-\frac{3}{2} \sigma)^4} \Theta \left(\frac{3}{2}\sigma-r\right),
\end{equation}
where $J$ is the spin-spin coupling strength, and $\epsilon \leq 10^{-4}$ is an arbitrarily small number to avoid division by zero. The form of $g(r)$ is chosen to both ensure continuity of the derivative $g'(r) \equiv dg(r)/dr$, for $r \leq 3\sigma/2$, and small values of $g(r)$ for $r \rightarrow 3\sigma/2$ \cite{Bhattacharya2025}. 

The linear and angular momenta are defined as:
\begin{equation}
\begin{aligned}
        \mathbf{p}_i &= m \mathbf{\dot{r}}_i + K \mathbf{S}_i, \qquad \omega_i = I \dot{\theta}_i,
\end{aligned}
\end{equation}
with the usual linear momentum now having a contribution from the spin degree of freedom.
Notably, this structure is formally analogous to the coupling of a charged particle to a magnetic vector potential, where the spin acts as an internal gauge field; a detailed derivation of this Lorentz force analogy is provided in Appendix~\ref{app:Lorentz_force}.

Coordinates are constrained to a periodic box of side length $L$, in two spatial dimensions ($\lbrace \mathbf{r} \rbrace \subseteq [-L/2,L/2]^2 \subseteq \mathbb{T}^2$). For each particle we define the orientation vectors
\[
\mathbf{n}_i=(\cos\theta_i, \sin\theta_i), \qquad 
\mathbf{n}_i^\perp=(-\sin\theta_i, \cos\theta_i),
\]
where $\theta_i \in [0,2\pi)$, so that the spin vector reads $\mathbf{S}_i = S \mathbf{n}_i$ and its time derivative is $\dot{\mathbf{S}}_i = S \dot{\theta}_i \mathbf{n}_i^\perp$. See Fig.~\ref{fig:model_and_snapshots}(a) for the schematic. Throughout, we set the spin magnitude $S=1$ for simplicity.

We next couple the system Eq.~\eqref{eq:Hamilton} to a thermal bath, at temperature $T$, \`{a} la Langevin. The overdamped dynamics read as
\begin{equation}\label{eq:OD_Langevins}
\begin{aligned}
\gamma_t \dot{\mathbf r}_i + K \dot{\theta}_i \mathbf n_i^\perp
&= \mathbf F_i(\{\mathbf r,\theta\}) + \boldsymbol\zeta_{i,r}(t), \\
\gamma_r \dot{\theta}_i - K \mathbf n_i^\perp \cdot \dot{\mathbf r}_i
&= {\tau}_i(\{\mathbf r,\theta\}) + \zeta_{i,\theta}(t),
\end{aligned}
\end{equation}
where $ \boldsymbol\zeta_{i,r}$ and $ \zeta_{i,\theta}$ are translational and rotational Gaussian white noises with variances $2\gamma_t k_B T$ and $2\gamma_r k_B T$, respectively, and the forces and torques are,
\begin{equation}
\begin{aligned}
\mathbf F_i(\{\mathbf r,\theta\})
&= - \sum_{j\neq i} \Big[ U_\mathrm{R}'(r_{ij})
- g'(r_{ij})\cos(\theta_{ji}) \Big]\hat{\mathbf r}_{ij}, \\
{\tau}_i(\{\mathbf r,\theta\})
&= \sum_{j\neq i} g(r_{ij})\sin(\theta_{ji}),
\end{aligned}
\end{equation}
where $\theta_{ji} \equiv \theta_j - \theta_i$, and $\hat{\mathbf r}_{ij}=(\mathbf{r}_i-\mathbf{r}_j)/|\mathbf{r}_i-\mathbf{r}_j|$. Equivalently, Eq.~\eqref{eq:OD_Langevins} can be written in block-matrix form as
\begin{equation}
\mathbf A_i(\theta_i)
\begin{pmatrix}
\dot{\mathbf r}_i \\[2pt] \dot{\theta}_i
\end{pmatrix}
=
\begin{pmatrix}
\mathbf F_i \\[2pt] \tau_i
\end{pmatrix}
+
\boldsymbol\zeta_i(t),
\end{equation}
with the friction matrix
\[
\mathbf A_i(\theta_i) =
\begin{pmatrix}
\gamma_t \mathbf I_2 & K \mathbf n_i^\perp \\
- K (\mathbf n_i^\perp)^\top & \gamma_r
\end{pmatrix},
\]
The diagonal entries $\gamma_t\mathbf I_2$, with $\mathbf{I}_2$ the 2$\times$2 identity matrix, and $\gamma_r$ are the ordinary translational and rotational Stokes drags. The off–diagonal blocks $K \mathbf n_i^\perp$ and $-K(\mathbf n_i^\perp)^\top$ encode the reversible spin–velocity coupling.
Let $\Delta=\gamma_t\gamma_r+K^2>0$ for invertibility. The explicit 2D overdamped Langevin equations for the translational and rotational degrees of freedom are then written as
\begin{equation}\label{eq:OD_Langevins_explicit}
\begin{aligned}
\begin{pmatrix}
\dot{\mathbf r}_i \\[2pt] \dot{\theta}_i
\end{pmatrix}
&=
\mathbf M_i(\theta_i)
\begin{pmatrix}
\mathbf F_i \\[2pt] \tau_i
\end{pmatrix}
+
\mathbf M_i(\theta_i) \boldsymbol\zeta_i(t),
\end{aligned}
\end{equation}
where the state-dependent mobility matrix $\mathbf M_i(\theta_i)\equiv\mathbf A_i(\theta_i)^{-1}$ reads
\begin{equation}\label{eq:Mobility_matrix}
\begin{aligned}
\mathbf M_i(\theta_i) &=
\begin{pmatrix}
\mathbf M_{rr} \quad \mathbf M_{r\theta} \\
 \mathbf M_{\theta r} \quad \mathbf M_{\theta \theta}
\end{pmatrix}\\ &=
\begin{pmatrix}
\frac{1}{\gamma_t} \left[\mathbf I_2 - \frac{K^2}{\Delta}\, \mathbf n_i^\perp(\mathbf n_i^\perp)^\top\right] & -\frac{K}{\Delta}\, \mathbf n_i^\perp \\
\frac{K}{\Delta}\, (\mathbf n_i^\perp)^\top & \frac{\gamma_t}{\Delta}
\end{pmatrix},
\end{aligned}
\end{equation}
with the noises collected as $\boldsymbol\zeta_i(t)=(\boldsymbol\zeta_{i,r}(t), \zeta_{i,\theta}(t))^\top$ which satisfy the fluctuation–dissipation relation (at the friction level)
\begin{equation}
\big\langle \boldsymbol\zeta_i(t)\, \boldsymbol\zeta_j(t')^\top \big\rangle
= 2 k_B T\,
\begin{pmatrix}
\gamma_t \mathbf I_2 & 0 \\[2pt]
0 & \gamma_r
\end{pmatrix}\;
\delta_{ij}\, \delta(t-t') \,.
\end{equation}

\subsection{Numerical Methods}

We integrate the dynamics \eqref{eq:OD_Langevins_explicit} using the Euler-Maruyama algorithm with a time step $\Delta t = 10^{-4}$ \cite{Note_drift}. Simulations are conducted in a two–dimensional box of size $L^2$ with periodic boundary conditions and at fixed packing fraction $\eta = N \pi (\sigma/2)^2 / (L^2)$. To systematically study the liquid-gas phase coexistence, we also run slab simulations where the box is made rectangular ($L_x = 3 L_y$) \cite{Julian2015,Siebert2018,Solon2018}. Particles are initially placed in a dense strip of width $L_x/2$ centered in the simulation box, with the remaining volume left empty. This setup minimizes nucleation barriers and facilitates the formation of stable liquid-gas interfaces perpendicular to the $x$-axis. For each set of parameters, simulations are performed for a total of $4 \times 10^7$ time steps. The initial $30\%$ of the trajectory is discarded as a relaxation period, as worked out from checking the saturation of observables such as density histograms as a function of time, with the remaining data used for analysis. Unless otherwise specified, all reported data are averaged over 5 independent simulation replicas.

\section{Results}

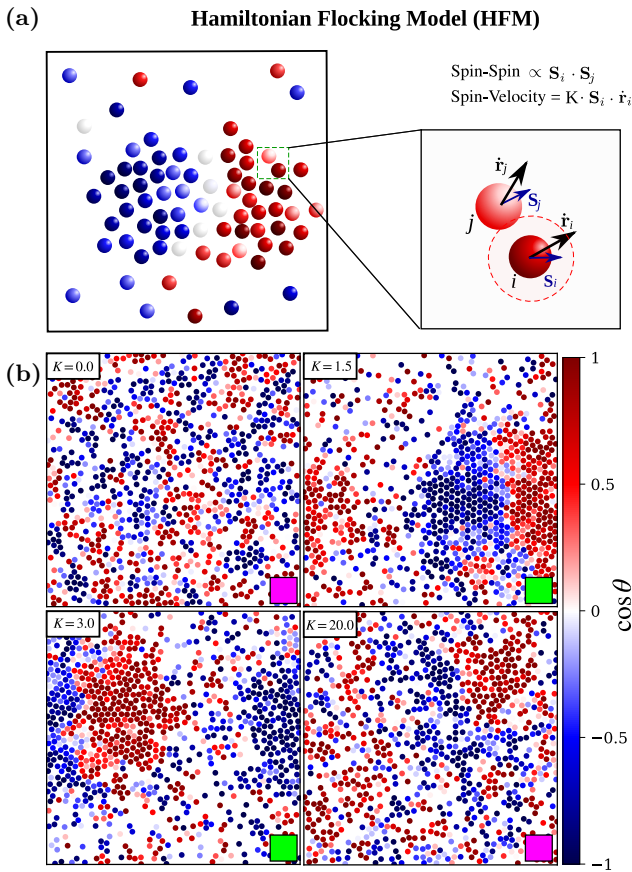
\begin{figure}[t]
  \centering
  
  \includegraphics[width=0.9\columnwidth]{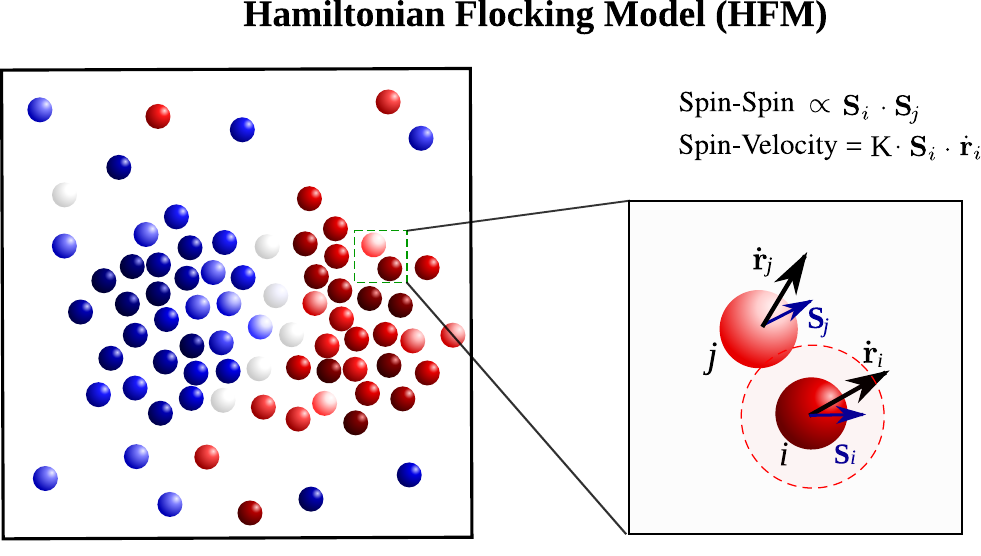}
  \label{fig:schematic} 
  \begin{picture}(0,0)
    \put(-240, 120){\textbf{(a)}} 
  \end{picture}

  \vspace{0.1cm} 

  \includegraphics[width=0.9\columnwidth]{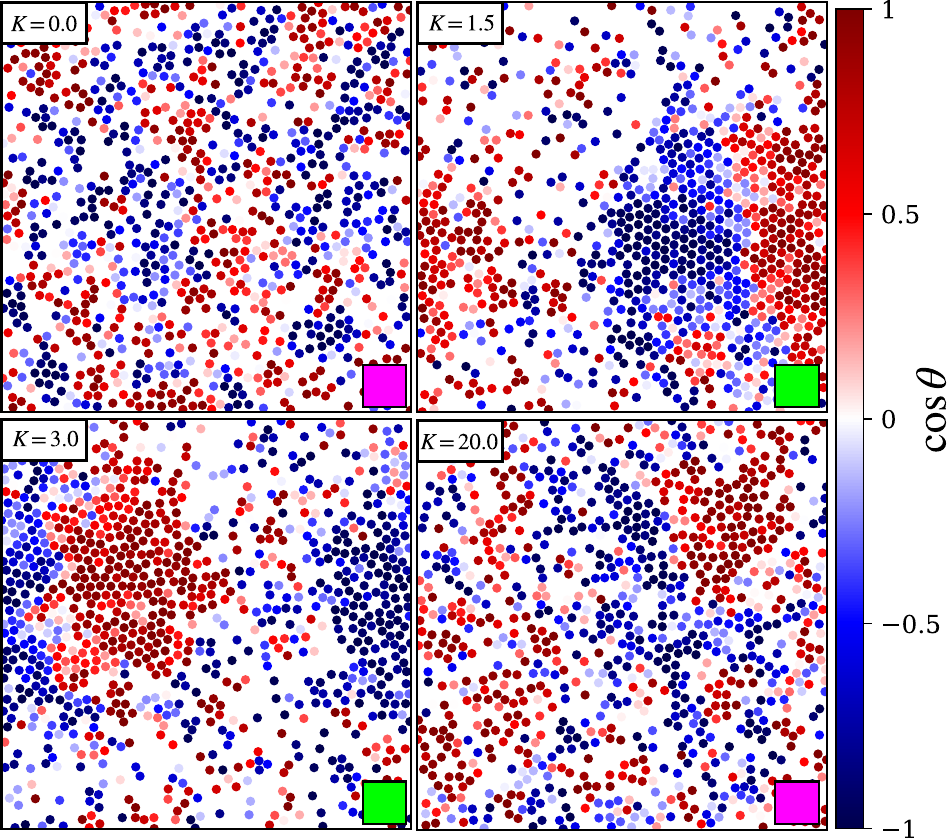}
  \label{fig:snapshots} 
  \begin{picture}(0,0)
    \put(-240, 185){\textbf{(b)}}
  \end{picture}

  \caption{ \textbf{Hamiltonian flocking model (HFM) and qualitative observation of reentrance.}
  \textbf{(a)} Pictorial description of the essential model ingredients defined in \eqref{eq:Hamilton}. Colors indicate orientation.
  \textbf{(b)} Steady-state snapshots at fixed $N=1000$, $\eta=0.32$, $T=0.6$, and $J=1$, for varying spin-velocity coupling strengths $K$ (top-left of snapshots).
  Particles are colored by their orientation $\cos\theta_i$. 
  Upon monotonically increasing $K$ the system exhibits reentrance: from a homogeneous phase (pink square) to clustered (green square) back to the homogeneous phase.  
  }
  \label{fig:model_and_snapshots}
\end{figure}

A striking feature of the HFM is the appearance of phase separation in the absence of an attractive pair potential (with explicit distance dependence), a behavior that has not been observed in equilibrium systems \cite{Casiulis2020} (see Fig.~\ref{fig:model_and_snapshots}). It is important to note, however, that the ferromagnetic spin–spin interaction in the Hamiltonian, Eq.~\eqref{eq:Hamilton}, namely $-g(r_{ij})\mathbf S_i\!\cdot\!\mathbf S_j$, can generate an effective spatial attraction between aligned spins and, under suitable conditions, induce FIPS~\cite{Casiulis_2019}. To disentangle this mechanism from the effects of the spin–velocity coupling, in the simulations below we choose a sufficiently high temperature $T$ such that, at the spin–velocity coupling strength $K=0$, the system remains in a homogeneous (gas) phase for the densities and couplings considered~\cite{Casiulis_2019}.

\textit{Reentrant phase behavior}:-Our key finding is in the clear demonstration of reentrant phase behavior in this model with the spin-velocity coupling ($K$) being the key control parameter. Our simulations show that as $K$ increases, the system displays reentrance: from a homogeneous gas to phase-separated coexistence, back to the homogeneous phase (Fig.~\ref{fig:model_and_snapshots}(b)). As the reentrance involves tunnelling through a clustered phase we aim to quantify this reentrant phase behavior using the differences in high- and low-density peaks of the local density distribution $\Delta\rho =\rho_{\mathrm{high}}-\rho_{\mathrm{low}}$ as the order parameter (see Appendix~\ref{app:density_extraction} for details on the density analysis). To observe reentrance, we hypothesize that $\Delta \rho(K)$ should be non-monotonic in $K$. 

\begin{figure}[t!]
  \centering
  \setlength{\tabcolsep}{0pt} 
  
  \begin{minipage}{\linewidth}
    \centering
    \includegraphics[width=0.999\linewidth, trim=0 0 0 0, clip]{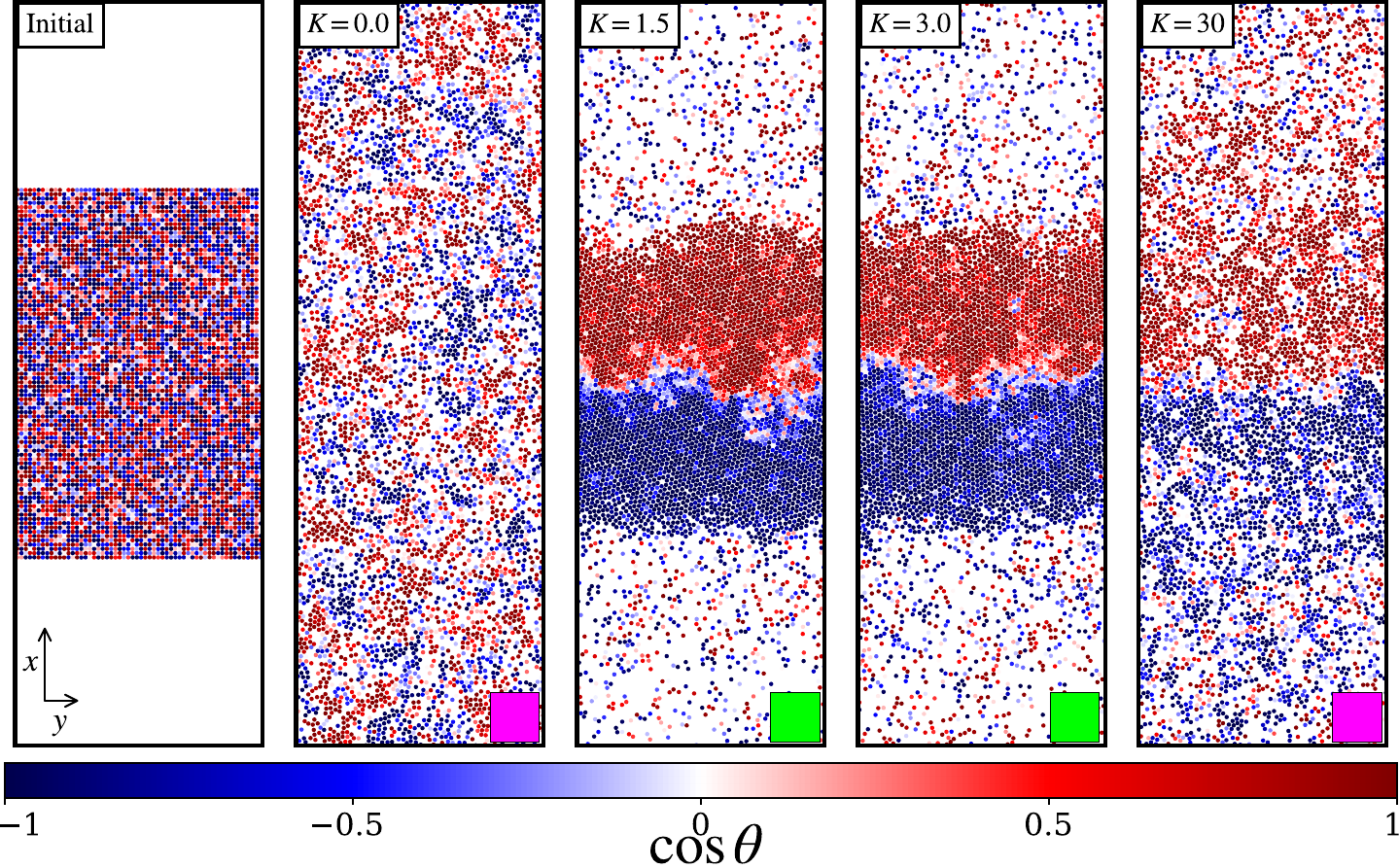}
    \put(-240, 135){\scriptsize \textbf{(a)}}
  \end{minipage}%

  \begin{minipage}{0.5\linewidth}
    \centering
    \includegraphics[width=\linewidth, trim=0 3 5 0, clip]{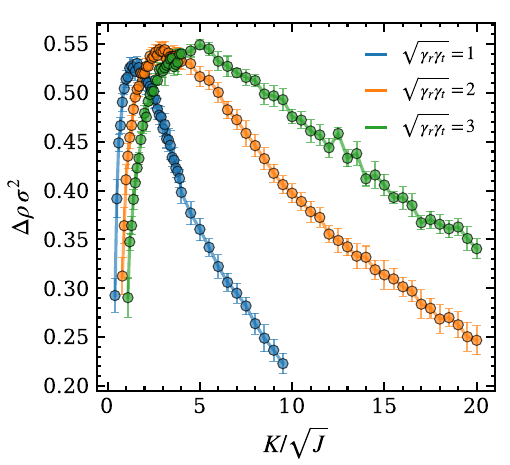}
    \put(-98, 101){\scriptsize \textbf{(b)}}
  \end{minipage}%
  \begin{minipage}{0.5\linewidth}
    \centering
    \includegraphics[width=\linewidth, trim=0 3 5 0, clip]{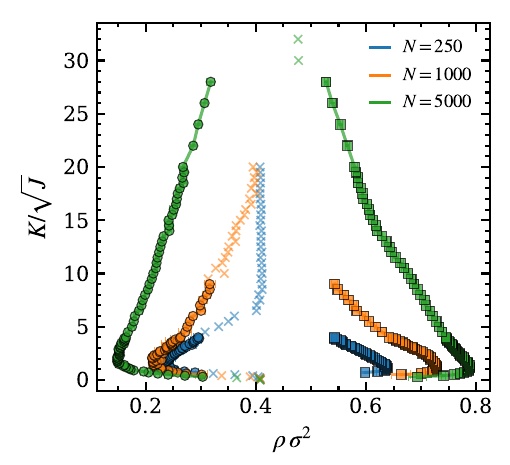}
    \put(-98, 101){\scriptsize \textbf{(c)}}
  \end{minipage}

  \begin{minipage}{0.5\linewidth}
    \centering
    \includegraphics[width=\linewidth, trim=0 0 5 5, clip]{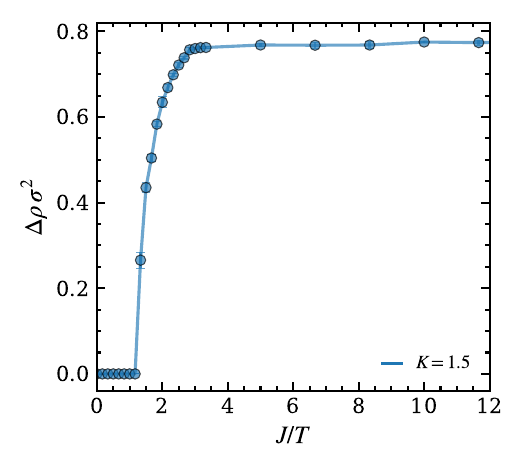}
    \put(-98, 102){\scriptsize \textbf{(d)}}
  \end{minipage}%
  \begin{minipage}{0.5\linewidth}
    \centering
    \includegraphics[width=\linewidth, trim=0 0 5 5, clip]{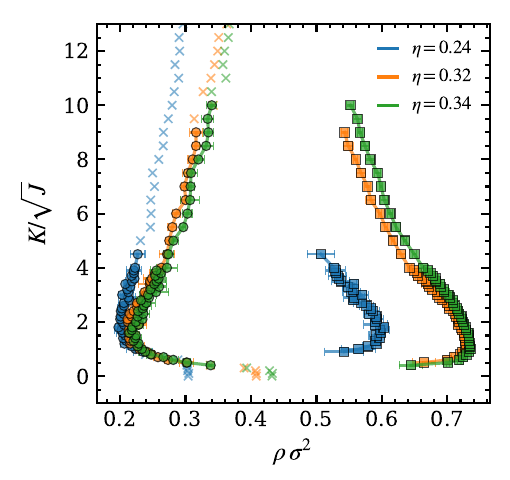}
    \put(-98, 104){\scriptsize \textbf{(e)}}
  \end{minipage}

   \caption{
 \textbf{Numerical characterization of reentrant phase behavior. }  
  \textbf{(a)} Representative snapshots of the slab simulations. From left to right: Initial configuration ($t=0$); steady states at $K=0, 1.5, 3, \text{ and } 30$. System size: $N=5000$ and $L_x = 3 L_y$.
  \textbf{(b)} Difference between high and low densities $\Delta\rho\sigma^2$ vs. scaled coupling $K/\sqrt{J}$ for varying $\sqrt{\gamma_t\gamma_r}$. The order parameter attains its maximum around $K = \sqrt{3\gamma_t\gamma_r}$. 
  \textbf{(c)} Phase coexistence diagram in the $K/\sqrt{J}$-$\rho$ plane for different $N$. Circles denote dilute phase, squares the dense phase, and gray crosses denote the single-phase.
  \textbf{(d)}  Order parameter $\Delta\rho \sigma^2$ as a function of the scaled ferromagnetic coupling $J/T$ at fixed $K=3/2$. The order parameter increases monotonically with $J/T$ and eventually saturates, confirming that stronger alignment interactions enhance phase separation.
  \textbf{(e)}  Phase diagram in the $K/\sqrt{J}$-$\rho$ plane for different global packing fractions $\eta \in \{0.24, 0.32, 0.34\}$. Solid lines with symbols represent the binodal coexistence densities, while gray crosses indicate homogeneous states. Unless otherwise indicated parameters used: $N=1000$, $T=0.6$, $\gamma_t=\gamma_r=1$, $\eta=0.32$, and $J=1$.  
  }
  \label{fig:combined_results}
\end{figure}

Next we quantify reentrance by performing slab simulations to facilitate the probing of phase separation and to remove the finite-size in one of the spatial dimensions (see Fig.~\ref{fig:combined_results}). As expected, our slab simulations recapitulate the observed reentrance behavior (Fig.~\ref{fig:combined_results}(a)). Indeed, upon monotonically increasing $K$ we find that $\Delta\rho(K)$ is non-monotonic in $K$, thus establishing true reentrance involving a transition from a gas to liquid-gas coexistence, and back again (see Fig.~\ref{fig:combined_results} (b)). We then wondered whether reentrance is robust to variations in dynamical parameters, system size, and packing fraction. First, we probed the properties of the viscous bath and examined the dependence of $\Delta \rho$ on the friction coefficients $\gamma_t \times \gamma_r$ (see Fig.~\ref{fig:combined_results}(b)). To highlight the competition between the spin–velocity coupling and the ferromagnetic alignment strength, we use the scaled variable $K/\sqrt{J}$.
Within the ranges explored, reentrance remains robust with changes in $\gamma_t \gamma_r$ shifting the location of the peak ($\max \Delta \rho(K)$) and also, though more subtly, the maximum value. We find that the order parameter attains its maximum around $K_{\mathrm{peak}} \approx \sqrt{3\gamma_t\gamma_r}$, which we rationalize further later in the manuscript. Second, to probe finite size effects we ran simulations of varying particle numbers $N \in \{250, 1000, 5000\}$, at fixed packing fraction, as shown in Fig.~\ref{fig:combined_results}(c). 
The character of both binodal lines remains qualitatively consistent as $N$ increases, suggesting that the observed reentrance is a genuine phase transition. We note that the maximum density contrast $\Delta \rho \sigma^2$ increases with system size. This scaling trend is consistent with recent findings regarding reentrant behavior in active Brownian particles exhibiting MIPS \cite{Yamamoto2025}.

To verify that phase separation is indeed driven by the interplay between alignment and coupling, we vary the ferromagnetic strength $J$ at fixed $K=3/2$. We show that the order parameter $\Delta \rho(J/T)$ exhibits a transition upon increasing $J$ where it is zero below a critical value $J/T < J^\ast/T \approx 1.2$, where alignment is destroyed by thermal noise, and the system remains homogeneous (see Fig.~\ref{fig:combined_results}(d)). Above the critical point, $\Delta \rho(J/T>J^\ast/T)$ rises rapidly and saturates, confirming that local ferromagnetic alignment is a prerequisite for the coupling-induced clustering and therefore, to some degree, controls reentrance \cite{Paoluzzi2024}. Finally, we show that changing the global packing fraction $0.24 < \eta < 0.34$ results in minor shifts in overall densities and does not impede reentrance, as shown in Fig.~\ref{fig:combined_results}(e). Collectively, these results establish that reentrant phase separation is a generic feature of the Hamiltonian flocking model.

\begin{figure*}[t]
  \centering

  \begin{minipage}{0.245\linewidth}
    \centering
    \includegraphics[width=\linewidth, trim=0 0 0 0, clip]{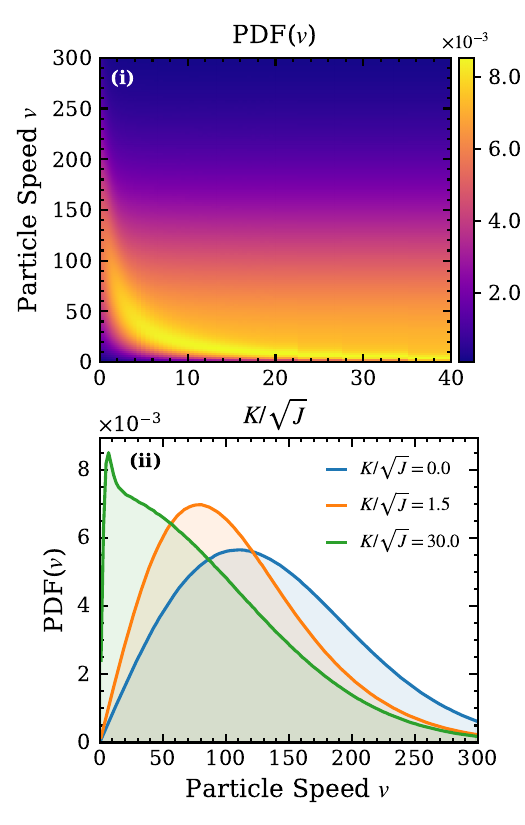}
    \put(-125, 195){\small \textbf{(a)}} 
  \end{minipage}
  \hfill 
  \begin{minipage}{0.245\linewidth}
    \centering
    \includegraphics[width=\linewidth, trim=10 0 0 0, clip]{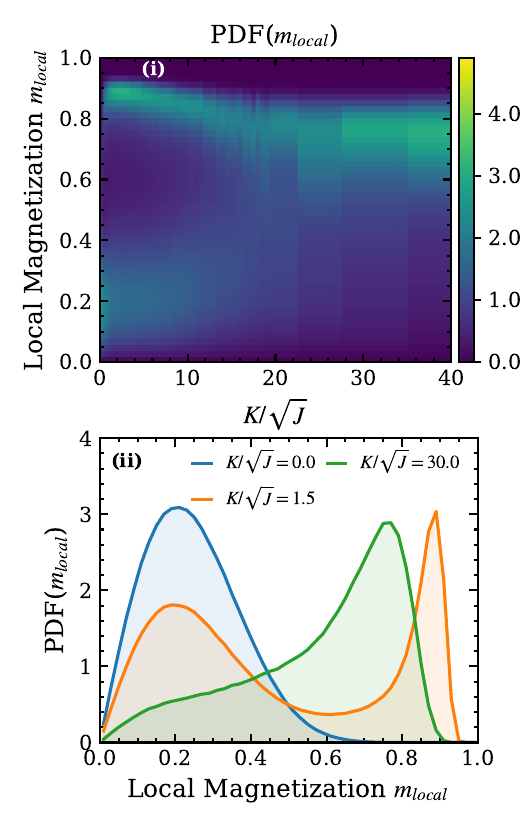}
    \put(-125, 195){\small \textbf{(b)}}
  \end{minipage}
  \hfill
  \begin{minipage}{0.245\linewidth}
    \centering
    \includegraphics[width=\linewidth, trim=0 0 0 0, clip]{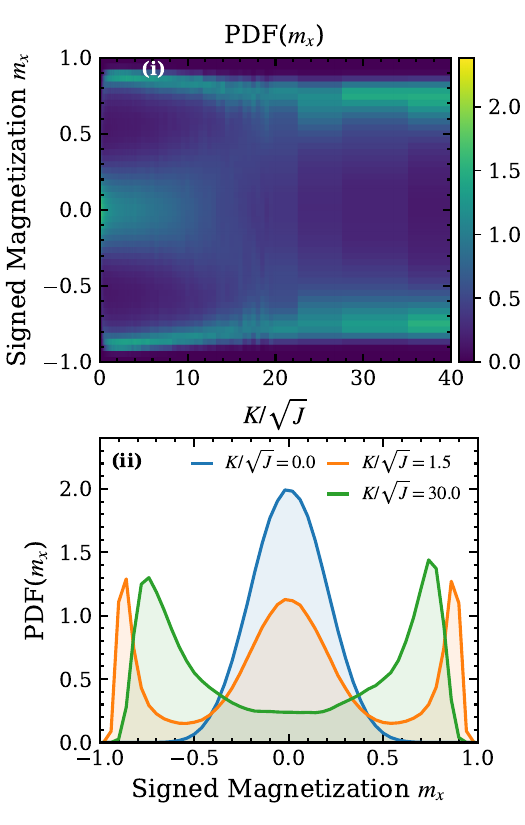}
    \put(-125, 195){\small \textbf{(c)}} 
  \end{minipage}
  \hfill 
  \begin{minipage}{0.245\linewidth}
    \centering
    \includegraphics[width=\linewidth, trim=2 0 0 0, clip]{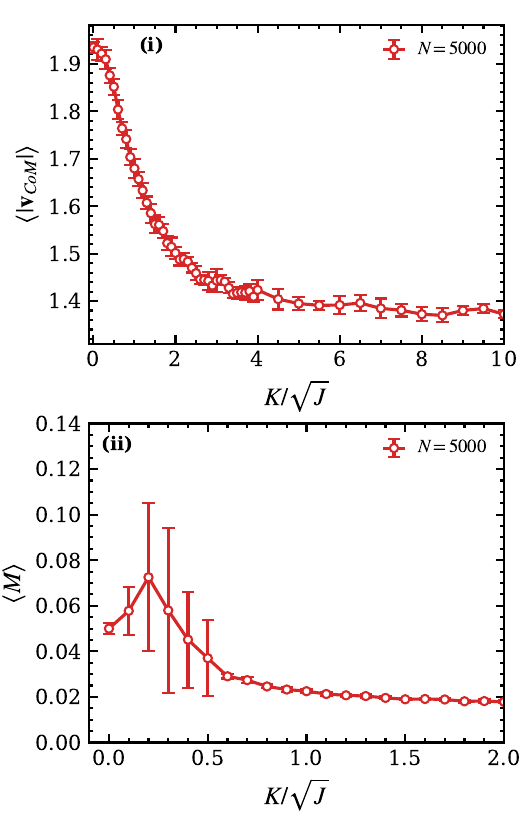}
    \put(-125, 195){\small \textbf{(d)}}
  \end{minipage}

  \caption{
   \textbf{Dynamical and structural properties across reentrance.} 
  This figure is organized into four columns corresponding to different physical quantities.
  \textbf{(a)} Particle speed analysis. \textbf{(i)} Heatmap of the particle speed distribution $P(v)$ vs scaled coupling $K/\sqrt{J}$. \textbf{(ii)} PDFs of particle speed $v$ for selected $K/\sqrt{J}$. 
  \textbf{(b)} Local magnetization analysis. \textbf{(i)} Heatmap of the local magnetization distribution  $P(m_{\text{local}})$ vs scaled coupling $K/\sqrt{J}$. \textbf{(ii)} PDFs of local magnetization for selected $K/\sqrt{J}$.
  \textbf{(c)} Local signed magnetization analysis. \textbf{(i)} Heatmap of the local signed magnetization distribution  $P(m_{x})$ vs scaled coupling $K/\sqrt{J}$. \textbf{(ii)} PDFs of local signed magnetization for selected $K/\sqrt{J}$. 
  \textbf{(d)} Macroscopic order parameters. \textbf{(i)} The mean center-of-mass velocity magnitude $\langle |\mathbf{v}_{\text{CoM}}| \rangle$ as a function of $K/\sqrt{J}$. \textbf{(ii)} The time-averaged global magnetization $\langle M \rangle$ as a function of $K/\sqrt{J}$, averaging over 10 simulation runs. All data are for $N=5000$, $T=0.6, \eta=0.32, J=1$, and $\gamma_t=\gamma_r=1$.
  }
  \label{fig:combined_micro_macro_N5000}
\end{figure*}

\textit{Properties across the reentrant transition}:-Having demonstrated the existence of reentrant phase behavior in the 2D overdamped Hamiltonian flocking model, we next probe the microscopic and macroscopic properties that characterize this transition (Fig.~\ref{fig:combined_micro_macro_N5000}). To understand the changes in dynamics we tracked the evolution of the particle speed distribution $P(v; K)$ with increasing coupling strength $K$ (Fig.~\ref{fig:combined_micro_macro_N5000}(a)). As shown in the heatmap Fig.~\ref{fig:combined_micro_macro_N5000}(a-i), and the representative PDFs Fig.~\ref{fig:combined_micro_macro_N5000}(a-ii), particles move more slowly at higher $K$. In the uncoupled limit ($K=0$), the system exhibits a standard Maxwellian distribution characteristic of an equilibrated gas. As $K$ increases the distribution deviates significantly, concentrating towards lower speeds. This slowing down at high coupling is a symptom of kinetic frustration, where the spin-velocity coupling likely restricts particle mobility.

After probing the dynamics, we next wondered how changes in $K$ affect the local structure (beyond that of particle positions) such as the local magnetization (see Fig.~\ref{fig:combined_micro_macro_N5000}(b) and also \cite{Bhattacharya2025}). We define the coarse-grained local magnetization magnitude $m_{\text{local}}(x) = | \sum_{i \in \text{bin}(x)} \mathbf{S}_i | / N_{\text{bin}(x)}$, calculated within the same spatial bins used for the density profile. The heatmap of the probability distribution $P(m_{\text{local}})$, in Fig.~\ref{fig:combined_micro_macro_N5000}(b-i) and selected profiles Fig.~\ref{fig:combined_micro_macro_N5000}(b-ii), reveal a clear topological transition in the probability landscape: as $K$ increases, the distribution transforms from unimodal--centered at a low finite value due to local finite-number fluctuations--to bimodal, indicative of the coexistence of a high-density ordered liquid and a low-density gas. 
At large $K$, $P(m_{\text{local}})$ reverts to a unimodal shape; however, as $m_{\text{local}}$ is a magnitude, this observation alone does not preclude the existence of long-wavelength polarity.

To resolve the underlying polarity, we consider the local signed magnetization $m_x(x) = (\sum_{i \in \text{bin}(x)} \cos \theta_i)/ N_{\text{bin}(x)}$ (see Figs.~\ref{fig:combined_micro_macro_N5000}(c-(i-ii))). For small $K$, the signed distribution $m_x(x)$ is unimodal and centred near zero, consistent with a homogeneous gas with no net local polarity. In the phase-separated regime, $m_x(x)$ becomes trimodal, with a central peak near $m_x(x) \approx 0$ (vapor phase) coexisting with two outer peaks at $m_x(x) \approx \pm 0.9$ (dense phase), indicating two oppositely polarized sub-domains in the dense cluster. At large $K$, the system is density-homogeneous. However, $P(m_x)$ transitions from trimodal to bimodal, with, now, only two peaks at $m_x(x)\approx \pm 0.75$. 
This reveals that in slab geometries with high spatial symmetry breaking, defined by an aspect ratio $L_x/L_y \gg 1$ (or $L_y/L_x \ll 1$), the high-$K$ homogeneous density regime supports long-lived bipolar spin domains that are separated by a stable, nearly flat domain wall (see Fig.~\ref{fig:combined_results}(a)). In contrast, square domains ($L_x/L_y = 1$) lack this spatial symmetry breaking and therefore relax to a homogeneous disordered state (see Appendix \ref{app:aspect_ratio_snapshots} for a comparison across different aspect ratios). 

We next examined the macroscopic, averaged over all particles and trajectories, versions of the dynamic and structural observables (see Fig.~\ref{fig:combined_micro_macro_N5000}(d)).
We show, in Fig.~\ref{fig:combined_micro_macro_N5000}(d-i), the mean center-of-mass velocity magnitude $\langle |\mathbf{v}_{\text{CoM}}| \rangle \equiv \langle| \sum_{i} \dot{\mathbf{r}}_i(t) | \rangle/N$ as a function of $K$.
Notably, this quantity decreases monotonically with increasing $K$ and is the clearest evidence that larger coupling hinders net transport. In contrast, the global orientational order exhibits a non-monotonic, reentrant-like dependence (see Fig.~\ref{fig:combined_micro_macro_N5000}(d-ii)).
This is quantified using the time-averaged global magnetization $\langle M \rangle$, defined via the instantaneous order parameter $M(t) = \frac{1}{N} \left| \sum_{i=1}^N \mathbf{S}_i(t) \right|$. The reported $\langle M \rangle$ is obtained by averaging $M(t)$ over the steady state and 10 independent simulation runs. Indeed, as shown in Fig.~\ref{fig:combined_micro_macro_N5000}(d-ii), $\langle M \rangle$ rises from a finite (bare) value to a maximal value at $K\approx0.2$ in the intermediate coupling regime. This signifies the formation of an ordered cluster before decaying back to near zero at large $K$.
This decoupling of transport speed (monotonic decay) and orientational order (reentrant peak) underscores the dual nature of the parameter $K$: it drives alignment at intermediate values but imposes kinetic frustration at large spin-velocity coupling.

\begin{figure*}[t]
  \centering
  \begin{minipage}{0.325\linewidth}
    \centering
    \includegraphics[width=\linewidth]{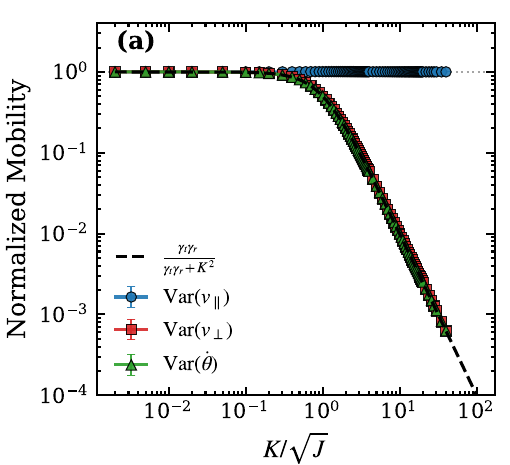}
  \end{minipage}
  \hfill
  \begin{minipage}{0.325\linewidth}
    \centering
    \includegraphics[width=\linewidth]{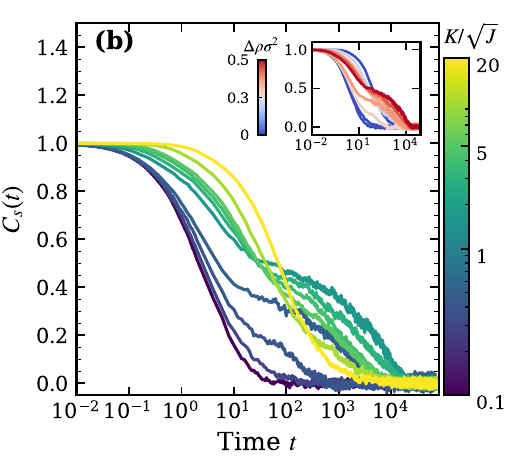}
  \end{minipage}
  \hfill
  \begin{minipage}{0.325\linewidth}
    \centering
    \includegraphics[width=\linewidth]{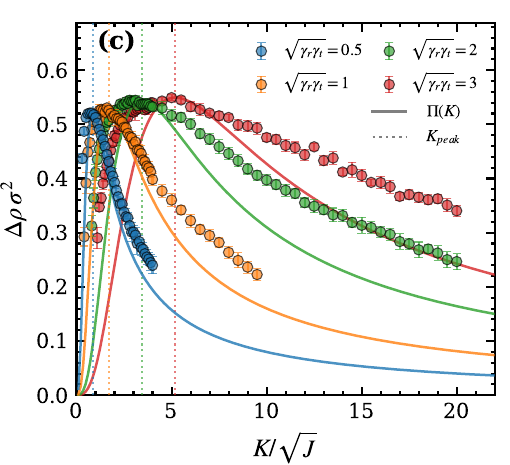}
  \end{minipage}

  \caption{ \textbf{Underlying reentrance mechanisms.}
  \textbf{(a)} Verification of kinetic frustration: Normalized mobility of longitudinal ($v_\parallel$, blue circles), transverse ($v_\perp$, red squares), and angular ($\dot{\theta}$, green triangles) degrees of freedom as a function of the scaled coupling $K/\sqrt{J}$. The simulations are performed with $N=1000$ particles at $J=1.0$ and $\gamma_t=\gamma_r=1$. The dashed black line represents the theoretical prediction derived from the Langevin dynamics (see Eq.~\ref{eq:variances}). At large coupling, the transverse and rotational variances decay as $K^{-2}$ (slope $-2$), while longitudinal motion remains unaffected, signifying a transition from isotropic diffusion to highly anisotropic diffusion.
  \textbf{(b)} The rotational autocorrelation function $C_s(t) = \langle \mathbf{S}_i(t) \cdot \mathbf{S}_i(0) \rangle$ vs. time $t$ for increasing scaled coupling strength $K/\sqrt{J}$. The inset presents the same data colored by the order parameter $\Delta\rho\sigma^2$. \textbf{(c)} Scaling ansatz of reentrance: Comparison of the density contrast $\Delta\rho\sigma^2$ (symbols) with the scaling ansatz $\Pi(K) \propto K^3/(\gamma_t\gamma_r + K^2)^2$ (solid curves, normalized to match the peak height of the simulation data) across different damping regimes. The vertical dotted lines indicate the theoretical peak position $K_{\mathrm{peak}} = \sqrt{3\gamma_t\gamma_r}$.}
  \label{fig:kinetic_frustration}
\end{figure*}

\textit{Mechanism of reentrance}:-Thus far, we have established reentrance in the HFM with the spin-velocity coupling $K$ as the control parameter and have shown how proxies for dynamics and structure indicate potential frustration. Next, we provide evidence to show that the reentrant phase behavior observed here arises from a competition between two $K-$coupling-induced effects: an effective drive that promotes clustering and a kinetic frustration that inhibits it. This competition manifests distinctly across three regimes of the spin-velocity coupling strength.

In the weak coupling regime ($K \ll K^*$, where $K^* = \sqrt{\gamma_t \gamma_r}$), thermal fluctuations dominate (see Appendix \ref{app:smallcoupling} for asymptotic scaling of the equations of motion in this regime). Although the ferromagnetic interaction $J$ induces local alignment, the spin-velocity coupling is insufficient to convert this orientational order into significant directed transport leading the system to behave as a disordered gas. As the coupling increases to the intermediate regime ($K \sim K^*$), see Appendix \ref{app:mediumcoupling} for the dynamics for this case, the system enters a Hamiltonian flocking phase. Here, the short-range ferromagnetic interaction aligns the spins of neighboring particles, while the spin-velocity coupling rectifies this alignment into directed motion. This mechanism generates an ``effective motility" analogous to the self-propulsion speed $v_0 = |\mathbf{v}|$ seen in standard active matter models.

Crucially, at large coupling ($K \gg K^*$), kinetic frustration dominates. In this regime, the effective mobility matrix becomes highly anisotropic. As indicated by the asymptotic analysis of the dynamics (see Appendix~\ref{app:largecoupling} for details), the transverse mobility scales as $\sim K^{-2}$, while the longitudinal mobility remains constant.
Physically, this implies that large $K$ causes the particles to effectively behave as 1D sliders confined to trajectories defined by their instantaneous spin.
Even when particles collide, the suppression of transverse diffusion prevents the geometric rearrangements and lateral sliding required to coalesce into dense clusters.
Furthermore, the rotational response to torques is similarly suppressed, locking the particle orientations.
This effective dimensional reduction—from 2D diffusive dynamics to 1D constrained motion—penalizes the formation of a 2D liquid phase, forcing the system to reenter a homogeneous, albeit kinetically arrested, gaseous state.

We validate this kinetic frustration mechanism by analyzing the particle velocity fluctuations. In the overdamped regime, the fluctuation-dissipation theorem dictates that the variance of the stochastic velocity components is determined by the diagonal elements of the mobility matrix, $\langle v_\alpha^2 \rangle \propto k_B T \mathbf{M}_{\alpha\alpha}$. By projecting the translational mobility tensor, Eq.~\eqref{eq:Mobility_matrix}, onto the instantaneous heading direction $\mathbf{n}_i$ and its orthogonal complement $\mathbf{n}_i^\perp$, we derive the explicit theoretical predictions for the longitudinal ($v_\parallel = \dot{\mathbf{r}}_i \cdot \mathbf{n}_i$), transverse ($v_\perp = \dot{\mathbf{r}}_i \cdot \mathbf{n}_i^\perp$), and rotational angular velocities:
\begin{equation}
\label{eq:variances}
\begin{aligned}
    \langle v_\parallel^2 \rangle &\propto \mathbf{n}_i^\top \mathbf{M}_{rr} \mathbf{n}_i = \frac{1}{\gamma_t}, \\
    \langle v_\perp^2 \rangle &\propto (\mathbf{n}_i^\perp)^\top \mathbf{M}_{rr} \mathbf{n}_i^\perp = \frac{\gamma_r}{\gamma_t\gamma_r + K^2}, \\
    \langle \dot{\theta}^2 \rangle &\propto M_{\theta\theta} = \frac{\gamma_t}{\gamma_t\gamma_r + K^2}.
\end{aligned}
\end{equation}
Upon normalization by their respective uncoupled values ($K=0$), both the transverse and rotational variances collapse onto a single scaling function, $(\gamma_t\gamma_r)/(\gamma_t\gamma_r + K^2)$.

In Fig.~\ref{fig:kinetic_frustration}(a) we compare the analytical predictions above with the measured variances from simulations. The data shows excellent agreement between these theoretical predictions and simulations across all coupling regimes. Importantly, while the longitudinal mobility remains constant ($\langle v_\parallel^2 \rangle \sim \text{const.}$), both the transverse and rotational variances exhibit a Lorentzian-like decay. In the limit of strong coupling ($K \gg \sqrt{\gamma_t\gamma_r}$), they follow the asymptotic scaling
\begin{equation}
      \langle v_\perp^2 \rangle \sim K^{-2}, \quad \langle \dot{\theta}^2 \rangle \sim K^{-2}.
\end{equation}
This quantitative confirmation proves that strong spin-velocity coupling acts as a kinetic constraint: it selectively suppresses the transverse and rotational degrees of freedom essential for rearrangement, effectively reducing the system to a collection of 1D sliders and thereby closing the phase separation window.

To further elucidate the dynamical nature of the high-coupling homogeneous state, we examine the temporal relaxation of the orientational degrees of freedom. In Fig.~\ref{fig:kinetic_frustration}(b), we plot the rotational autocorrelation function, $C_s(t) = \langle \mathbf{S}_i(t) \cdot \mathbf{S}_i(0) \rangle$, for increasing spin-velocity coupling strengths.
In the weak coupling regime, $C_s(t)$ decays quickly, characteristic of simple diffusive motion. As $K$ increases, a distinct intermediate plateau emerges, a hallmark of kinetic arrest~\cite{Berthier_2011}. 
At larger K ($K \gg \sqrt{\gamma_t\gamma_r}$), the two-step relaxation collapses into a single, very slow decay and the intermediate plateau visibly weakens or disappears. 
The inset recolors the same curves by the order parameter $\Delta\rho\sigma^2$, highlighting that the plateau is most pronounced near the peak of $\Delta\rho\sigma^2$ (intermediate $K$) and fades both in the low- and high-$K$ homogeneous phases.

\textit{Analogy with motility-induced phase separation}:-Despite the conservative nature of the HFM, the observed reentrant phase separation bears a striking resemblance to the reentrance observed in repulsive active fluids \cite{Su2023}. In repulsive active Brownian particles (RABPs), the drive towards clustering is often quantified by the swim pressure, $\Pi_{\mathrm{swim}} \propto v_0^2 \tau_r$, where $v_0$ is the self-propulsion speed and $\tau_r \propto 1/D_r$ is the rotational relaxation time (or persistence time) \cite{Takatori_2014,Solon_2015}. 
Phase separation occurs when this active drive overcomes the stabilizing thermal diffusion, and where the density-dependent instantaneous speed $v(\rho)$ is monotonically decreasing in $\rho$ \cite{Solon_2015}.

In the HFM, there is no intrinsic self-propulsion. Instead, particle flux is induced by the spin-velocity coupling. To adapt the swim pressure framework, we construct a scaling function $\Pi(K)$ by identifying the effective drift velocity, relaxation time, and coupling efficiency from the microscopic Langevin dynamics, Eq.~\eqref{eq:OD_Langevins_explicit}. The effective drift velocity $v_{\text{eff}}$ arises from the conversion of torque to linear motion scaling with the cross-mobility coefficient magnitude, $v_{\text{eff}} \propto|M_{r\theta}| \propto K/\Delta$ where $\Delta = \gamma_t\gamma_r + K^2$. The rotational relaxation time, determined by the inverse rotational diffusion, scales as $\tau_r \propto \Delta/\gamma_t$.  Crucially, unlike ABPs where motility is intrinsic, the HFM requires a coupling process to rectify torque-induced displacements into a sustained flux. This coupling efficiency is naturally proportional to the coupling strength normalized by the total friction, $\epsilon(K) \propto |M_{\theta r}| \propto K/\Delta$. Thus, $\epsilon(K)$ quantifies how much effective pressure is generated from the alignment torque to counteract the diffusive dissolution of the cluster.

Multiplying the base ``activity'' $v^2\times\tau_r \sim (K/\Delta)^2\times \Delta$ with the coupling efficiency $\epsilon(K) \propto K/\Delta$ results in a scaling ansatz for the effective coupling dependent swim pressure:
\begin{equation}
    \Pi(K) \sim \frac{K^3}{(\gamma_t\gamma_r + K^2)^2}.
    \label{eq:scaling_ansatz}
\end{equation}
This simple expression captures the non-monotonicity of $\Delta \rho$ as seen in Fig.~\ref{fig:combined_results}(b): for $K \ll \sqrt{\gamma_t \gamma_r}$, $\Pi(K) \sim K^3$ whereas for $K \gg \sqrt{\gamma_t \gamma_r}$, $\Pi(K) \sim K^{-1}$. Increasing $K$ in the intermediate-coupling regime enhances an activity-like drive and promotes clustering, while further increasing $K$ suppresses both transverse and rotational mobilities through the growth of $\Delta$, producing kinetic frustration and reentrance.
The peak of the transition corresponds to the maximum of Eq.~\eqref{eq:scaling_ansatz}. Solving $d \Pi(K) / dK = 0$ yields the condition $K_{\text{peak}} = \sqrt{3\gamma_t\gamma_r}$, which is in excellent agreement with our simulations (Fig.~\ref{fig:kinetic_frustration}c). We note that the deviation between the simulation data and the theoretical curves is expected, as the scaling ansatz $\Pi(K)$ is a minimal phenomenological proxy rather than a microscopic solution to the many-body Fokker-Planck equation. Specifically, the theory treats the system at a mean-field level, neglecting crowding effects inside dense clusters, and ignoring spatial correlations at the liquid-gas interface \cite{Takatori_2014,Solon_2015}.

Despite these gross simplifications, the ansatz successfully captures the non-monotonic nature of the transition and, crucially, accurately predicts the shift of the optimal coupling $K_{\mathrm{peak}}$ with friction parameters, confirming that the reentrance mechanism is indeed governed by the competition between a coupling-induced effective drive and mobility-limited kinetic frustration.

\section{conclusions and outlook}

In summary, we have demonstrated that reentrant phase separation, reminiscent of that observed in active matter, naturally emerges in a Hamiltonian flocking model. Our findings challenge the view that motility-induced clustering and its suppression require non-conservative energy injection. By analyzing the microscopic Langevin dynamics, we identified the underlying mechanism as a competition between an {effective drive}, which rectifies spin alignment into directed motion, and a kinetic frustration, which suppresses transverse mobility at strong coupling. Specifically, we showed that the system undergoes an effective dimensional reduction that cages particle orientations in the large coupling limit. This kinetic constraint prevents the geometric rearrangement necessary for cluster nucleation, thereby restoring a homogeneous phase. A scaling ansatz derived from this mechanism quantitatively captures the reentrant phase boundaries observed in simulations.

The reentrant behavior driven by the spin-velocity coupling strength shown here resembles, but is ultimately different from, the reentrant glass transition observed in dense active matter \cite{Paoluzzi2024}. Here, the suppression of transverse diffusion at large coupling drives a reentrance of the homogeneous phase from a clustering regime, as opposed to the situation in \cite{Paoluzzi2024} where a transition between an active fluid and moving glass states occurs, with strong collective migration suppressing local particle rearrangements.

Our work contributes to a unified understanding of reentrant behavior in both passive and active systems, highlighting the role of kinetic constraints in shaping binodal lines. The type of reentrance shown here results in clear non-monotonicity in both the dilute and dense binodal lines, which we deem strong reentrance, whereas in other systems only one binodal line is affected \cite{Burekovic2026}. This may be rationalized through the underlying frustration mechanism: here, the dimensional reduction occurs through large (self) spin-velocity coupling which does not require high densities and hence why we also see non-monotonicity in the dilute binodal line. An outstanding question is how to classify reentrant behaviour in both passive and active systems in terms of both frustration mechanism and how they affect the binodal lines \cite{Thomas2011}.

By generating phase behavior resembling that as seen in active systems, within a conservative Hamiltonian framework, our work underscores the surprising analogies between driven amorphous solids and dense active matter \cite{Puneet_2026}. Indeed, our work raises the question of how sharp are the boundaries between truly active systems and those that are externally driven \cite{Fodor2016}? The connecting of the HFM to a model with a Lorentz force Lagrangian implies that some active systems may resemble conservative ones that are subject to external driving with fields resolved at the level of individual constituents.

Future theoretical directions include placing \eqref{eq:scaling_ansatz} on a firm mathematical basis, most likely through a continuum mean-field approach \cite{Solon_2015}. Effort in this direction will likely also include further analysis of Hamiltonian ``active'' models \cite{Casiulis2020,Fieguth2022} in relation to standard non-conservative models of active matter. An interesting question is how reentrance affects the efficient control/driving of collective phase transitions. For instance, the physics outlined here reveals new strategies for controlling self-assembly through tunable kinetic frustration. Another avenue for further research is in exploring whether reentrance, as shown here, is borne out in topology (such as percolation) \cite{Levis2014,Evans2024} and geometry \cite{Davis2025}.

Importantly, our results are amenable to experimental testing in systems with spin-velocity coupling, such as active granular matter~\cite{Deseigne_2012,Weber_2013}, Quincke rollers~\cite{Bricard_2013}, and chiral fluids exhibiting odd viscosity~\cite{Banerjee_2017,Soni_2019}.

\begin{acknowledgments}
L.C. is funded by a Wellcome Accelerator Award (Davis 311948/Z/24/Z). L.K.D. is funded by a Flora Philip Fellowship at the University of Edinburgh.
Authors acknowledge the use of the Edinburgh Compute and Data Facility (ECDF).
\end{acknowledgments}

\bibliography{refs}

\appendix

\section{Analogy with the Lorentz Force}
\label{app:Lorentz_force}

The Lagrangian of the model reads \cite{Bhattacharya2025}:

\begin{equation}
\begin{aligned}
        \mathcal{L} &:= \mathcal{A} - \mathcal{V}, \\
        \mathcal{A}&= \sum_{i=1}^N \left(\frac{m \dot{\mathbf{{r}}}^2_i}{2} + \frac{I \dot{\theta}^2_i}{2} + K \mathbf{S}_i \cdot \dot{\mathbf{{r}}}_i \right), \\
        \mathcal{V} &=  \sum_{i<j}^N\left( U_\text{R}(r_{ij}) - g(r_{ij})\mathbf{S}_i \cdot \mathbf{S}_j  \right),
        \label{eq:HFMLagrangian}
\end{aligned}
\end{equation}
with the Hamiltonian being retrieved under a simple Legendre transformation
\begin{equation}
\mathcal{H} = \sum_{i=1}^N \left( \left( \frac{\partial \mathcal{L}}{\partial \dot{\mathbf{r}}_i} \right) \dot{\mathbf{r}}_i + \left( \frac{\partial \mathcal{L}}{\partial \dot{{\theta}}_i} \right) \dot{{\theta}}_i \right) - \mathcal{L},\\
\end{equation}
In Eq.~\eqref{eq:HFMLagrangian} the spin-velocity coupling term is reminiscent of a vector potential-velocity coupling term arising from a Lorentz force. For the model under consideration here, one can write an analogous (on the constituent level) Lorentz force as:
\begin{equation}
    \mathbf{F}_i = K \mathbf{v}_i \times \mathbf{B}_i,
\end{equation}
where $\mathbf{v}_i \equiv \dot{\mathbf{r}_i}$ and with the individual ``magnetic field'', $\mathbf{B}_i$, defined as:
\begin{equation}
    \mathbf{B}_i = \nabla_i \times \mathbf{S}(\mathbf{r}),
\end{equation}
with $\mathbf{S}(\mathbf{r})$ a microscopic spin field
\begin{equation}
    \mathbf{S}(\mathbf{r}) = \sum_{i=1}^N S_i \delta(\mathbf{r}-\mathbf{r}_i).
\end{equation}

One is then tempted to write down a heuristic ``electric field'', $\mathbf{E}$, which--when assuming a typical electromagnetic form--reads as:
\begin{equation}
    \mathbf{E} = -\nabla \varphi(\mathbf{r}) + \partial_t \mathbf{S}.
\end{equation}
The scalar field $\varphi(\mathbf{r})$ has the form:
\begin{equation}
    \varphi(\mathbf{r}) \propto \mathbf{S}(r) \cdot \mathbf{S}(r),
\end{equation}
then Eq.~\eqref{eq:HFMLagrangian} takes a form proximal to a Lorentz force Lagrangian:
\begin{equation}
\begin{aligned}
         \mathcal{L} &= \mathcal{L}_\text{LF} + \sum_{i=1}^N \frac{I \dot{\theta}^2_i}{2} - \sum_{i<j}^N U_\text{R}(r_{ij}), \\
         \mathcal{L}_\text{LF} &= \frac{m}{2} \sum_{i=1}^N \mathbf{v}_i^2 - K \varphi + K \sum_{i=1}^N \mathbf{v}_i \cdot \mathbf{S}(\mathbf{r}_i),
\end{aligned}
\end{equation}
where $\mathcal{L}_\text{LF}$ is the classical form of the Lagrangian for charged particles moving under (here microscopic) magnetic fields. This suggests that the Hamiltonian Flocking Model just described can be viewed as a driven system, but with the \textit{external} driving localized on the individual constituents.

\section{Extraction of Binodal Lines and Density Analysis}
\label{app:density_extraction}

To accurately determine the coexisting gas ($\rho_{\text{gas}}$) and liquid ($\rho_{\text{liq}}$) densities defining the binodal lines in the phase diagram (Fig.~\ref{fig:combined_results}), we employ a statistical analysis of the local density fluctuations rather than a simple spatial averaging. This approach is particularly robust against the strong interfacial fluctuations.

\subsection{Local Density Sampling}
For each set of parameters $(K, \rho_0)$, the system is initialized in a slab configuration, see Fig.~\ref{fig:combined_results}(a), and simulated for a total of $T_{\text{total}}$ time steps. We discard the initial $30\%$ of the trajectory as the relaxation period to ensure the system has reached a steady state. 
For the remaining production run, we compute the instantaneous 1D density profile $\rho(x, t)$ along the longitudinal direction ($L_x$). The simulation box is divided into $N_b = L_x / \Delta x$ spatial bins of width $\Delta x \approx 2.0\sigma$. The local density in the $i$-th bin at time $t$ is given by $\rho_i(t) = n_i(t) / (L_y \Delta x)$, where $n_i(t)$ is the number of particles in the bin.

\subsection{Probability Distribution Construction}
Instead of fitting the spatial profile $\rho(x)$ to a sigmoid function (e.g., $\tanh$), which can be biased by capillary waves, we construct the probability distribution function (PDF), $P(\rho)$, of the local densities. We collect the set of all local density values $\{\rho_i(t)\}$ sampled from all spatial bins over the entire steady-state time window.

To mitigate discretization artifacts from histograms and to filter out high-frequency shot noise which is common in finite-size systems, we estimate $P(\rho)$ using Gaussian Kernel Density Estimation (KDE):
\begin{equation}
    P(\rho) = \frac{1}{N_{\text{sample}} h} \sum_{j=1}^{N_{\text{sample}}} \mathcal{K}\left( \frac{\rho - \rho_j}{h} \right),
\end{equation}
where $\mathcal{K}$ is the standard normal kernel. We select a smoothing bandwidth factor $h$ (set to 0.2 in our implementation) that is sufficiently large to smooth out spurious peaks arising from finite-number fluctuations in the homogeneous phase, yet small enough to resolve the bimodal structure of phase separation.

\subsection{Peak Identification and Robustness Criteria}
The coexistence densities are identified as the local maxima of the smoothed PDF $P(\rho)$. If $P(\rho)$ exhibits two distinct peaks at densities $\rho_1$ and $\rho_2$ (with $\rho_1 < \rho_2$), we check the peak separation $\Delta \rho = \rho_2 - \rho_1$. To distinguish genuine phase separation from transient fluctuations in a homogeneous fluid, we impose a threshold criterion $\Delta \rho > \rho_{\text{thresh}}$ (set to $0.2$ in our implementation). If satisfied, we assign $\rho_{\text{gas}} = \rho_1$ and $\rho_{\text{liq}} = \rho_2$. If $P(\rho)$ is unimodal, or if the detected peaks are too close ($\Delta \rho \le \rho_{\text{thresh}}$), the system is classified as homogeneous (single-phase), and the density is recorded as the location of the global maximum.

\section{Asymptotic Analysis of Mobility Regimes}
\label{app:asymptotic_analysis}

Recall that the translational velocity $\dot{\mathbf r}_i$ (first row of Eq.~\eqref{eq:OD_Langevins_explicit}) is given by:
\begin{equation}\label{eq:r_dot}
\begin{split}
\dot{\mathbf r}_i &= \left(\frac{1}{\gamma_t} \mathbf{n}_i \mathbf{n}_i^\top + \frac{\gamma_r}{\Delta} \mathbf{n}_i^\perp (\mathbf{n}_i^\perp)^\top\right) \mathbf{F}_i 
+ \left(-\frac{K}{\Delta} \mathbf{n}_i^\perp\right) \tau_i \\
&\quad + \mathbf{M}_{rr} \boldsymbol\zeta_{i,r} + \mathbf{M}_{r\theta} \zeta_{i,\theta}.
\end{split}
\end{equation}

The rotational velocity $\dot{\theta}_i$ (second row of Eq.~\eqref{eq:OD_Langevins_explicit}) is given by:
\begin{equation}\label{eq:theta_dot}
\begin{split}
\dot{\theta}_i &= \frac{K}{\Delta} (\mathbf{n}_i^\perp)^\top \mathbf{F}_i 
+ \frac{\gamma_t}{\Delta} \tau_i + \mathbf{M}_{\theta r} \boldsymbol\zeta_{i,r} + M_{\theta\theta} \zeta_{i,\theta}
\end{split}
\end{equation}
where $\Delta \equiv \gamma_t \gamma_r + K^2$, and orientation basis $\mathbf{n} = (\cos \theta, \sin \theta)$ and $\mathbf{n}^\perp = (-\sin \theta, \cos \theta)$. Define $K^{*}\equiv\sqrt{\gamma_t \gamma_r}$.

\subsection{Small Coupling Regime ($K \ll K^*$), up to $K^{2}$ order.}
\label{app:smallcoupling}
In the regime of weak spin-velocity coupling, 
\[
\frac{\gamma_r}{\Delta} = \frac{1}{\gamma_t} \frac{1}{1 + K^2/(\gamma_t \gamma_r)} \simeq \frac{1}{\gamma_t} \left( 1 - \frac{K^2}{\gamma_t \gamma_r} \right),
\]
\[
\frac{K}{\Delta} \simeq \frac{K}{\gamma_t \gamma_r}, \quad \frac{\gamma_t}{\Delta} \simeq \frac{1}{\gamma_r} \left( 1 - \frac{K^2}{\gamma_t \gamma_r} \right).
\]

We have:
\begin{equation}
\begin{split}
\dot{\mathbf r}_i &= \left(\frac{1}{\gamma_t} \mathbf{n}_i \mathbf{n}_i^\top + \frac{1}{\gamma_t} \left( 1 - \frac{K^2}{\gamma_t \gamma_r} \right) \mathbf{n}_i^\perp (\mathbf{n}_i^\perp)^\top\right) \mathbf{F}_i \\
&-\frac{K}{\gamma_t \gamma_r} \mathbf{n}_i^\perp \tau_i + \mathbf{M}_{rr} \boldsymbol\zeta_{i,r} + \mathbf{M}_{r\theta} \zeta_{i,\theta},
\end{split}
\end{equation}

\begin{equation} 
\begin{split}
\dot{\theta}_i &= \frac{K}{\gamma_t \gamma_r} (\mathbf{n}_i^\perp)^\top \mathbf{F}_i 
+ \frac{1}{\gamma_r} \left( 1 - \frac{K^2}{\gamma_t \gamma_r} \right) \tau_i \\
&\quad + \mathbf{M}_{\theta r} \boldsymbol\zeta_{i,r} + M_{\theta\theta} \zeta_{i,\theta}.
\end{split}
\end{equation}

In the limit where the spin-velocity coupling is negligible ($K \to 0$), we have $\Delta \simeq \gamma_t \gamma_r$. Consequently, the off-diagonal coupling terms vanish ($K/\Delta \to 0$), and the diagonal mobility coefficients simplify to the standard inverse friction:
\[
\frac{\gamma_r}{\Delta} \simeq \frac{1}{\gamma_t}, \quad \frac{\gamma_t}{\Delta} \simeq \frac{1}{\gamma_r}.
\]
Crucially, the translational mobility becomes isotropic:
\[
\mathbf{M}_{rr} \simeq \frac{1}{\gamma_t} \mathbf{n}_i \mathbf{n}_i^\top + \frac{1}{\gamma_t} \mathbf{n}_i^\perp (\mathbf{n}_i^\perp)^\top = \frac{1}{\gamma_t} \mathbf{I}_2.
\]
The equations of motion decouple into standard overdamped Langevin equations for translation and rotation:
\begin{equation}
\begin{aligned}
\dot{\mathbf r}_{i} &= \frac{1}{\gamma_t}\,\mathbf F_i + \frac{1}{\gamma_t}\,\boldsymbol\zeta_{i,r}, \\
\dot{\theta}_{i} &= \frac{1}{\gamma_r}\,\tau_i + \frac{1}{\gamma_r}\,\zeta_{i,\theta}.
\end{aligned}
\end{equation}
This recovers the equilibrium dynamics of passive particles with spin, where velocity and orientation are statistically independent.

\subsection{Medium Coupling Regime ($K \sim  K^*$).}
\label{app:mediumcoupling}

At the critical coupling strength $K^* = \sqrt{\gamma_t \gamma_r}$, the determinant becomes $\Delta = 2\gamma_t\gamma_r$. The magnitude of the cross-coupling reaches its maximum, $|K/\Delta| = (2\sqrt{\gamma_t\gamma_r})^{-1}$. The diagonal coefficients become:
\[
\frac{\gamma_r}{\Delta} = \frac{1}{2\gamma_t}, \quad \frac{\gamma_t}{\Delta} = \frac{1}{2\gamma_r}.
\]
Substituting these into the dynamics, we obtain the specific equations of motion for the maximally coupled regime:
\begin{equation}
\begin{aligned}
\dot{\mathbf r}_i &= \left( \frac{1}{\gamma_t} \mathbf{n}_i \mathbf{n}_i^\top + \frac{1}{2\gamma_t} \mathbf{n}_i^\perp (\mathbf{n}_i^\perp)^\top \right) \mathbf{F}_i \\
&\quad - \frac{1}{2\sqrt{\gamma_t\gamma_r}} \mathbf{n}_i^\perp \tau_i +  \mathbf{M}_{rr} \boldsymbol\zeta_{i,r} + \mathbf{M}_{r\theta} \zeta_{i,\theta},
\end{aligned}
\end{equation}
\begin{equation}
\begin{aligned}
\dot{\theta}_i &= \frac{1}{2\sqrt{\gamma_t\gamma_r}} (\mathbf{n}_i^\perp)^\top \mathbf{F}_i + \frac{1}{2\gamma_r} \tau_i + \mathbf{M}_{\theta r} \boldsymbol\zeta_{i,r} + M_{\theta\theta} \zeta_{i,\theta}.
\end{aligned}
\end{equation}
Note that the translational mobility becomes anisotropic: the mobility along the heading direction $\mathbf{n}_i$ is $1/\gamma_t$, while the mobility perpendicular to the heading $\mathbf{n}_i^\perp$ is suppressed to $1/(2\gamma_t)$.

\subsection{Large Coupling Regime ($K \gg  K^*$).}
\label{app:largecoupling}
In the regime of strong spin-velocity coupling, the determinant is dominated by the coupling term, $\Delta \simeq K^2$. The mobility coefficients scale as follows:
\[
\frac{\gamma_r}{\Delta} \simeq \frac{\gamma_r}{K^2}, \quad \frac{\gamma_t}{\Delta} \simeq \frac{\gamma_t}{K^2}, \quad \frac{K}{\Delta} \simeq \frac{1}{K}.
\]
The translational mobility matrix $\mathbf{M}_{rr}$ becomes rank-one (reduced dimensionality):
\[
\mathbf{M}_{rr} \simeq \frac{1}{\gamma_t} \mathbf{n}_i \mathbf{n}_i^\top + \mathcal{O}\left(\frac{1}{K^2}\right).
\]
The leading-order equations of motion become:
\begin{equation} 
\begin{split}
\dot{\mathbf r}_i &= \left(\frac{1}{\gamma_t} \mathbf{n}_i \mathbf{n}_i^\top + \frac{\gamma_r}{K^2} \mathbf{n}_i^\perp (\mathbf{n}_i^\perp)^\top\right) \mathbf{F}_i 
+ \left(-\frac{1}{K} \mathbf{n}_i^\perp\right) \tau_i \\
&\quad + \mathbf{M}_{rr} \boldsymbol\zeta_{i,r} + \mathbf{M}_{r\theta} \zeta_{i,\theta},
\end{split}
\end{equation}

\begin{equation} 
\begin{split}
\dot{\theta}_i &= \frac{1}{K} (\mathbf{n}_i^\perp)^\top \mathbf{F}_i 
+ \frac{\gamma_t}{K^2} \tau_i + \mathbf{M}_{\theta r} \boldsymbol\zeta_{i,r} + M_{\theta\theta} \zeta_{i,\theta}.
\end{split}
\end{equation}

In the asymptotic limit of infinite coupling ($K \rightarrow \infty$), all terms scaling with $1/K$ and $1/K^2$ vanish. The mobility matrix $\mathbf{M}$ becomes singular ($M_{\theta \theta} \rightarrow 0$). The dynamics reduce to:
\begin{equation}
\dot{\mathbf r}_i = \left(\frac{1}{\gamma_t} \mathbf{n}_i \mathbf{n}_i^\top\right) \mathbf{F}_i =\frac{1}{\gamma_t} (\mathbf{n}_i \cdot \mathbf{F}_i) \mathbf{n}_i, \qquad \dot{\theta}_i = 0.
\end{equation}
In this limit, the angular degree of freedom is completely frozen ($\dot{\theta}=0$). The particle loses all ability to rotate or move sideways. It behaves as a 1D slider constrained to move strictly along a fixed line defined by its initial orientation, responding only to the projection of external forces along this axis. This gives a transition from a 2D flocking to a gas of effectively non-interacting (in terms of alignment) 1D sliders.

\section{Robustness of the Homogeneous Density Regime across Aspect Ratios}
\label{app:aspect_ratio_snapshots}

\begin{figure}[t]
  \centering
  \includegraphics[width=0.9\columnwidth]{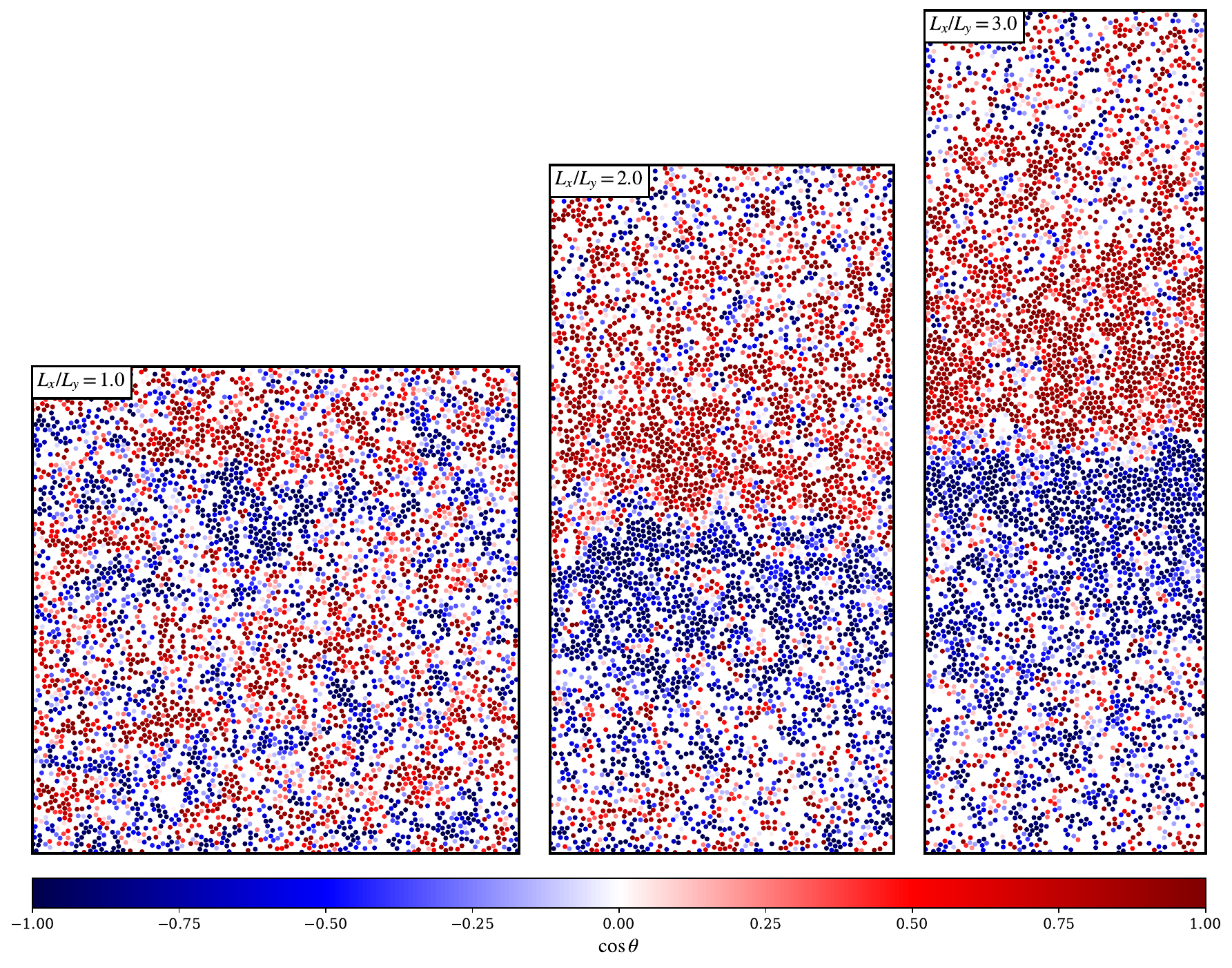}
  \caption{Steady-state snapshots across different aspect ratios.
  Results are shown for $L_x/L_y = 1, 2,$ and $3$ with fixed parameters  $N=5000$, $K=30$ , $T=0.6$ , $\eta=0.32$, $J=1$, and $\gamma_t=\gamma_r=1$. Increasing $L_x/L_y$ stabilizes a nearly flat spin domain wall, leading to long-lived bipolar spin domains at large $K$ while the density remains essentially homogeneous.}
  \label{fig:aspect_ratios}
\end{figure}

In this section, we examine the influence of the simulation box geometry on the steady-state configurations of the system at large $K$. Fig.~\ref{fig:aspect_ratios} presents snapshots for aspect ratios $L_x/L_y = 1, 2,$ and $3$ at a high coupling strength ($K=30$). 

Our primary observation is that the density distribution remains spatially homogeneous across all investigated geometries. This confirms that the reentrance into a density homogeneous phase at large $K$, as discussed in the main text, is a robust bulk property of the model.

In contrast, the spatial topology of the spin degrees of freedom is sensitive to the box shape. In slab geometries ($L_x/L_y=2$ and $3$), the elongated axis provides a preferred direction that stabilizes long-lived bipolar spin domains. Conversely, in the square domain ($L_x/L_y=1$), the absence of a dominant length scale prevents this geometric stabilization of the domain walls. In this case, the system lacks the longitudinal constraints required to support the bipolar state and instead relaxes toward a globally isotropic configuration. 

These results demonstrate that while the specific spatial topology of the spin texture is dictated by the box geometry, the reentrant transition to a density-homogeneous state-characterized by the suppression of cluster formation at high $K$-is independent of the aspect ratio.

\end{document}